\documentclass{elsart}

\usepackage{graphicx}
\usepackage{epsfig}
\usepackage{amsmath}
\usepackage{amssymb}
\usepackage{comment}
\usepackage{url}

\newcommand{\tens}[1]{\mathbf{#1}} 
\newcommand{\pd}{\partial}
\DeclareMathOperator{\Wi}{Wi}
\DeclareMathOperator{\tr}{tr}

\newcommand{\att}{\tens{A}}
\newcommand{\axx}{A_{xx}}
\newcommand{\axy}{A_{xy}}
\newcommand{\ayy}{A_{yy}}

\newcommand{\ttt}{\tens{T}}
\newcommand{\txx}{T_{xx}}
\newcommand{\txy}{T_{xy}}
\newcommand{\tyy}{T_{yy}}

\newcommand{\itt}{\tens{I}}

\begin{document}

  \begin{frontmatter}

    \title{Scaling of singular structures in extensional flow of
           dilute polymer solutions}
    \author[Leiden]{Paul Becherer\corauthref{cor1}},
      \ead{becherer@lorentz.leidenuniv.nl}
    \author[Leiden,Edinburgh]{Alexander N. Morozov} and
    \author[Leiden]{Wim van Saarloos}
    \corauth[cor1]{Corresponding author.}
    \address[Leiden]{Instituut-Lorentz for Theoretical Physics,
                     Universiteit Leiden,
                     Postbus 9506,
                     NL-2300 RA Leiden,
                     The Netherlands
                     }
    \address[Edinburgh]{School of Physics,
                        University of Edinburgh,
                        JCMB,
                        King's Buildings,
                        Mayfield Road,
                        Edinburgh EH9 3JZ,
                        United Kingdom
                      }
    \begin{abstract}
      Recently singular solutions have been discovered in purely elongational flows of visco-elastic fluids. We surmise that these solutions are the mathematical structures underlying the so-called birefringent strands seen experimentally. In order to facilitate future experimental studies of these we derive a number of asymptotic results for the scaling of the width and extension of the near-singular structures in the FENE-P model for polymers of finite extensibility.
       \end{abstract}

  \end{frontmatter}

\section{Introduction}
\begin{figure}
  \begin{center}
    \begin{tabular}{ccc}
      \includegraphics[width=0.2\textwidth]{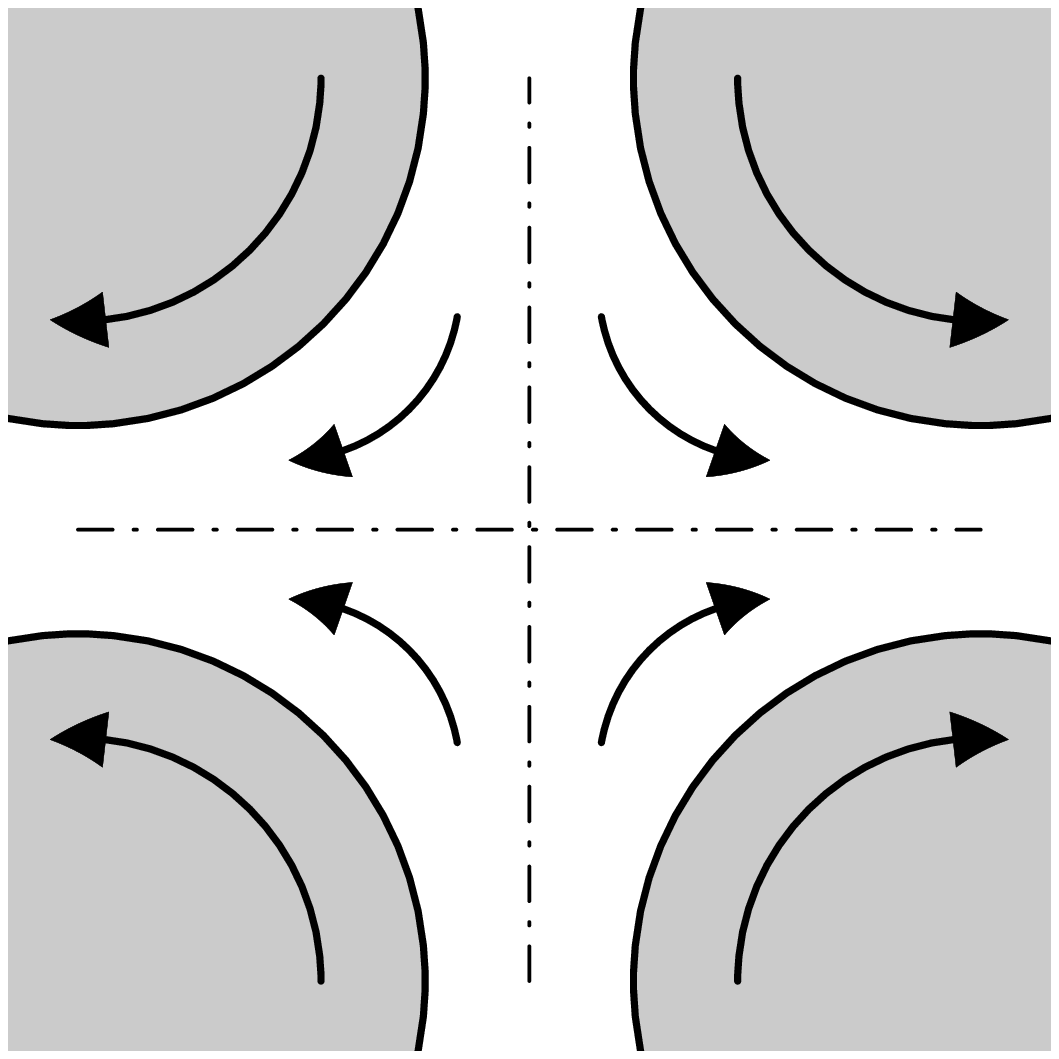} &
      \hspace{2cm} &
      \includegraphics[width=0.2\textwidth]{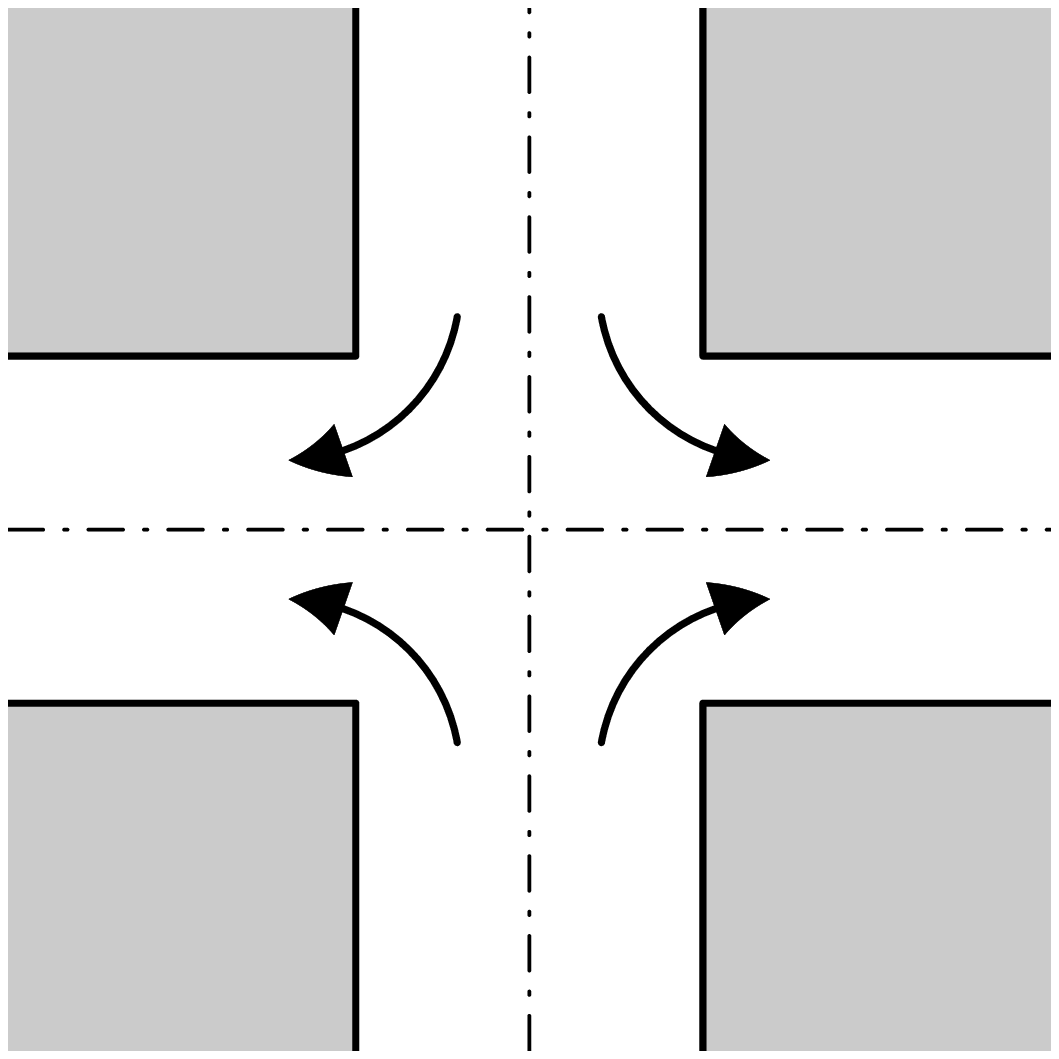} \\
      (a) & \hspace{2cm} & (b) 
    \end{tabular}
    \caption{(a) The four-roll mill (b) Cross-channel flow.
\label{fig:int_stag_flow}}
  \end{center}
\end{figure}

Recently, there has been renewed interest in the properties of extensional flows of dilute
polymer solutions, in particular in a class of flows known as
\emph{internal stagnation point flows}, such as the four roll mill
flow and the cross-slot or cross channel flows shown in
Fig.~\ref{fig:int_stag_flow}. Theoretical modelling of such flows has proven to be particularly challenging, and despite extensive experimental and theoretical
investigation, some aspects of these flows remain to be elucidated.
For example, Arratia \emph{et al.}~\cite{arratia2006} have recently
found a bifurcation to asymmetric stationary flow patterns in a cross-channel flow of a dilute polymer solution, followed by a secondary instability to time-dependent asymmetric flows.  In spite of its simple appearance, this instability is not yet 
fully understood. Conversely, instabilities have been found numerically
that have no clear experimental counterpart (for example, Harris
and Rallison~\cite{harris1993,harris1994} and Xi and
Graham~\cite{xi2007}).

In traditional investigations of elongational flows, it has often
been assumed, either implicitly or explicitly, that the base flow
solution is the classical solution \cite{bird1} of the Oldroyd-B
continuum equation in which the stresses are constant in space.
However, Rallison and Hinch~\cite{rallison1988} already mention
singular solutions of the UCM equations that are strongly peaked at the
centre line along the outflow direction, and Renardy~\cite{renardy2006}
recently pointed out that these singular solutions are also relevant
for $\Wi < 1/2$, where they do not actually diverge, but are still
singular. Although such solutions are
not easy to probe numerically, recent work by Thomases and
Shelley~\cite{thomases2006} shows that these solutions do emerge
spontaneously in high accuracy numerical simulations of a model problem
for elongational flows. One reason for this may be that, as we will
discuss, the constant stress solutions typically do not satisfy the
physical boundary conditions at the edges of the flow region. A
natural question that therefore emerges is whether these singular
solutions are the mathematical counterpart of the so-called
birefringent strands that have been found
experimentally~\cite{mackley1975,crowley1976,fuller1980} and
numerically~\cite{harlen1992,feng1997,remmelgas1999}. Such strands
are thin regions of highly extended polymers along the central
outgoing streamlines. A strand can modify the flow,
leading to a ``dip'' in the observed velocity profile of the
exit flow~\cite{rabin1986,harlen1990}, and it seems reasonable
to assume that they affect any instabilities that might occur.

The singular solutions mentioned by Renardy \cite{renardy2006}
arise in models like the UCM model which essentially assume that
the polymers are infinitely extensible. A second important question
that emerges from these studies is therefore how these singular
solutions are modified when we consider finitely extensible polymers,
as in the FENE-P model \cite{bird2}. The first step in this direction 
was made by Renardy \cite{renardy2006}, who showed that for the
Giesekus model, which limits the growth of elongational stresses,
stress profiles remain smooth while stress gradients can diverge.

The aim of this article therefore is to study the scaling of the main
characteristics (e.g., the width) of these singular solutions strands
in a simple way, and to analyse the modifications due to the finite
extensibility. Our analysis is done for the ideal case of purely
elongational flow, which is a good approximation near the stagnation
point. By confining the analysis to this simple case we are able to
derive a number of explicit asymptotic scaling results which we hope
will facilitate making the connection between the (almost) singular
solutions and the birefringent strands. 

Our results are consistent with results for steady flow 
by Renardy~\cite{renardy2006}, Thomases and Shelley~\cite{thomases2006}
and Xi and Graham~\cite{xi2007}. However, since our approach is
essentially one-dimensional, we can obtain a better numerical
resolution, and it may be hoped that some calculations that are not
(yet) feasible in a full two-dimensional approach, such as eigenvalue
calculations for stability analysis, may be performed within the
framework we present here. It should be mentioned, however, that
the current experimental resolution~\cite{odell2006} is substantially
worse than the numerical resolution, even for full 2-D numerics.

The layout of this paper is the following. In Section 2 we summarize
the equations for extensional flow of a UCM fluid, and the structure
of the singular solutions discovered recently. After analysing
extensional flow of a FENE-P fluid in Section 3, we derive in
Section 4 various asymptotic results for this case. We end the paper
with a brief discussion of the robustness of our results.

\section{Purely extensional flow of a UCM fluid}
Putting the stagnation point at $(x, y) = (0, 0)$, a purely extensional
flow field is given by
\begin{equation}
  \vec{v} = (v_x, v_y) = \dot{\epsilon} (x, -y),
  \label{elong_flow}
\end{equation}
where $\dot{\epsilon}$ is the {\em elongation rate}. 
This flow field satisfies incompressibility,
$\vec{\nabla} \cdot \vec{v} = 0$. In a nondimensionalized formulation
of the UCM model, the flow field becomes
\begin{equation}
  \vec{v} = (x, -y),
  \label{elong_flow_dimless}
\end{equation}
and the constitutive equation for steady flow is~\cite{bird1}
\begin{equation}
\ttt
+ \Wi \left[ 
        (\vec{v}\cdot\vec{\nabla}) \ttt
      - (\vec{\nabla} \vec{v})^T \cdot \ttt
      - \ttt\cdot(\vec{\nabla} \vec{v})
      \right]
=
\vec{\nabla}\vec{v} + (\vec{\nabla}\vec{v})^T \label{Tequation}.
\end{equation}
Here $\ttt$ is the stress tensor, $\vec{v}$ the velocity, and
$\Wi$ the \emph{Weissenberg number}, which for purely extensional
flow we define as
\begin{equation}
\Wi = \dot{\epsilon}\lambda,
\end{equation}
with $\lambda$ the relaxation time of the polymers. The momentum
conservation equation for creeping flow is
\begin{equation}
  \vec{\nabla} \cdot \ttt - \vec{\nabla} p = 0.
\end{equation}

Solutions for steady flow are found by inserting the pure
extensional flow into the constitutive equation:
\begin{equation}
  \begin{aligned}[]
    \txx + \Wi \left(
                    x \frac{\pd \txx}{\pd x}
                  - y \frac{\pd \txx}{\pd y}
                  - 2 \txx
               \right) &= 2, \\
    \txy + \Wi \left(
                    x \frac{\pd \txy}{\pd x}
                  - y \frac{\pd \txy}{\pd y}
               \right) &= 0, \\
    \tyy + \Wi \left(
                    x \frac{\pd \tyy}{\pd x}
                  - y \frac{\pd \tyy}{\pd y}
                  + 2 \tyy
               \right) &= -2. 
  \end{aligned}
\end{equation}
If we assume a spatially uniform stress field, these equations reduce
to
\begin{equation}
  \txx = \frac{2}{1 - 2\Wi};
  \quad \txy = 0;
  \quad \tyy = \frac{-2}{1 + 2\Wi}.
  \label{eqn:unif_sol_ucm}
\end{equation}

It is well-known that for $\Wi \geq 1/2$, this solution diverges
and becomes unphysical~\cite{bird1}. If we no longer require that
the stresses be constant in space, the stress fields lose smoothness
even below $\Wi = 1/2$, as was pointed out by
Renardy~\cite{renardy2006}. Around and above this
value of $\Wi$, there are large stress gradients around $y = 0$,
reminiscent of a birefringent strand. We shall derive these solutions,
essentially following Renardy's presentation.

\begin{figure}
  \begin{center}
    \includegraphics[width=0.7\textwidth]{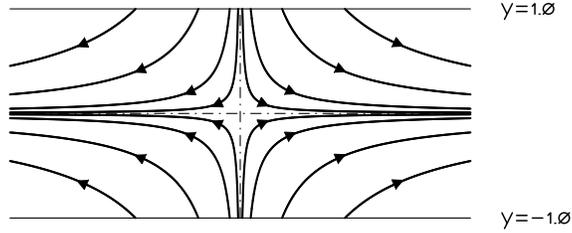} 
    \caption{Uniform extensional flow on an infinite strip.
\label{fig:ext_strip}}
  \end{center}
\end{figure}

We choose as flow domain the strip $(x, y) \in \Rset \times [-1, 1]$.
On this domain we assume pure extensional flow, see
Fig.~\ref{fig:ext_strip}. We assume that the
stresses depend only on $y$:
\begin{equation}
  \txx \equiv \txx(y); \quad \txy \equiv \txy(y);
  \quad \tyy \equiv \tyy(y) = p(y),
\end{equation}
such that momentum conservation is obeyed. We impose boundary
conditions for the normal stresses on the ``inflow''
boundaries~\cite{renardy1988}:
\begin{equation}
  \txx(\pm 1) = \xi;
  \quad \tyy(\pm 1) = \eta .
\end{equation}

The equations reduce to
\begin{equation}
  \begin{aligned}[]
    &\txx(y) + \Wi \left(-y \txx'(y) - 2 \txx(y) \right) &=& 2, \\
    &\txy(y) + \Wi \left(-y \txy'(y) \right) &=& 0, \\
    &\tyy(y) + \Wi \left(-y \tyy'(y) + 2 \tyy(y) \right) &=& -2.
  \end{aligned}
  \label{eqn:sol_ucm}
\end{equation}
The solution can be given in closed form\footnote{A similar solution
was given in~\cite{renardy2006}. The UCM equations reported there,
however, seem to be written for uniaxial extensional flow
$\vec{v} = (1,-\frac{1}{2},-\frac{1}{2})$ rather than for planar
extension $\vec{v} = (1,-1)$ as claimed in~\cite{renardy2006} and as
in the present paper. This explains the difference between
Eq.~(\ref{eqn:ucm_sol}) here and Eq.~(9) from~\cite{renardy2006}.}:
\begin{equation}
  \begin{aligned}[]
    \txx = & \frac{2}{1-2\Wi}
            + \left(\xi - \frac{2}{1-2\Wi}\right)|y|^{1/\Wi-2} \\
    \txy = & \,\,0\\
    \tyy = & \frac{-2}{1+2\Wi}
            + \left(\eta + \frac{2}{1+2\Wi}\right)|y|^{1/\Wi+2} \\
  \end{aligned}
\label{eqn:ucm_sol}
\end{equation}
The first part of the $xx$ and $yy$ stress components is the same as
for the uniform solution~(\ref{eqn:unif_sol_ucm}). The second part has
to be added to make the solution consistent with the imposed boundary
conditions.
Due to the fractional exponents in this part, none of these solutions
is smooth,
(except when $1 / \Wi$ is an integer). This becomes important when
approximating these functions using spectral methods.

As Eq.~(\ref{eqn:ucm_sol}) clearly shows, the classical constant stress
solution (\ref{eqn:unif_sol_ucm}) only exists for very special
boundary conditions. Indeed, the non-smooth terms in Eq.~(\ref{eqn:ucm_sol}) are present at every Weissenberg number unless
one chooses $\xi=2/(1-2\Wi)$ and $\eta=-2/(1+2\Wi)$.
As $\Wi$ approaches $1/2$, these terms create a narrow region of
large extensional stress, qualitatively similar to a birefringent
strand, as was already observed by Rallison and
Hinch~\cite{rallison1988}.

More interestingly, the purely elongational flow,
Eq.~(\ref{elong_flow}) does not support any boundary value
for the shear stress $T_{xy}$ other than $T_{xy}=0$ at
the boundary~\cite{renardy1988,hulsen1988}. Forcing $T_{xy} \neq 0$
at the inflow would
inevitably modify the velocity profile, Eq.~(\ref{elong_flow}): in
addition to the purely extensional flow field, it would acquire a
shear component. The importance of this modification will be
discussed in Section 5 where we address the relevance of our results
to the stability of experimental realizations of stagnation-point
flows \cite{arratia2006}.

Note also that even in the range where the uniform solution clearly
breaks down ($\Wi \geq 0.5$), the non-uniform
solution~(\ref{eqn:ucm_sol}) is well-defined. However, the stress on
the central outgoing streamline diverges, and for $\Wi \geq 1$, the
total elastic energy, which is proportional to the integral of
$\tr \tens{T}$, also diverges.

\section{Extensional flow in a FENE-P model}
The FENE-P model avoids the blow-up of the extensional stress by
implementing a nonlinear force law for the polymer
molecules~\cite{bird2}. The solvent is usually treated explicitly.
We shall ignore the solvent viscosity, thus formulating an extension
of the UCM model rather than the Oldroyd-B model.

We shall formulate the FENE-P model in terms of the
\emph{conformation tensor}
\begin{equation}
  \att = \langle \vec{R}\vec{R}\rangle,
\end{equation}
where $\vec{R}$ is the end-to-end vector connecting two beads of
a single dumbbell. The brackets denote an ensemble average.

In our nondimensionalization, the stress is given in terms of the
conformation tensor as
\begin{equation}
  \ttt = \frac{1}{\Wi}
                \left(
                  \frac{\att}{1-\tr{\att}/L^2} 
                  - \frac{\itt}{1-2/L^2}
                \right).
\label{eqn:stress_conformation}
\end{equation}
Here, $\itt$ is the unit tensor and $L^2$ is the maximal
value of the trace of the conformation tensor, that is, $L$ is the
maximal extension of the dumbbells, relative to their equilibrium
extension. The $2$ appears in the rightmost denominator because in
two spatial dimensions this is $\tr \itt$.

The constitutive equation for steady flow then is~\cite{thomases2006}
\begin{equation}
  \Wi \left[
          (\vec{v}\cdot\vec{\nabla}) \att
           - (\vec{\nabla} \vec{v})^T \cdot \att
           - \att \cdot (\vec{\nabla} \vec{v})
        \right]
        =
        -\left(
           \frac{\att}{1-\tr \att / L^2}
           - \frac{\itt}{1 - 2 / L^2}
         \right).
\label{eqn:fene_p_ce}
\end{equation}
Note that Eqs.~(\ref{eqn:stress_conformation}) and~(\ref{eqn:fene_p_ce}) differ
from the ``classical'' FENE-P model in that they are restricted to two dimensions
and assume the other components of the conformation tensor 
($A_{xz}$, $A_{yz}$, $A_{zz}$) to be zero. In the two-dimensional flow Eq.~(\ref{elong_flow_dimless}),
the classical FENE-P model~\cite{bird1} would have $A_{zz}\neq0$. However, this approximation
bears no influence on the asymptotic result for the width of the birefringent strand found in Section 4.

The momentum balance is nonlinear in the conformation tensor:
\begin{equation}
  \frac{1}{\Wi}\left(
  \frac{\vec{\nabla}\cdot\att}{1-\tr \att/L^2}
  +\frac{\att\cdot\vec{\nabla}(\tr \att)}%
        {L^2(1 - \tr \att /L^2)^2}
        \right)
  -\vec{\nabla}p
  = 0.
\label{eqn:fene_p_mb}
\end{equation}
Again, if we assume that the conformation tensor (and, hence, the
stress tensor) depends only on $y$, this expression simplifies, and
we find one equation for $\axx$ and $\ayy$, and an
equation for $\axy$ in terms of $\axx$ and $\ayy$:
\begin{equation}
  \frac{1}{1 - \tr \att/L^2}
    \left(
      \frac{\pd \ayy}{\pd y} 
      + \frac{\ayy}{1 - \tr \att / L^2}
        \frac{\pd \tr \att / L^2}{\pd y}
    \right)
    - \Wi\frac{\pd p}{\pd y}
    = 0.
\end{equation}
and
\begin{equation}
  \frac{1}{1 - \tr \att / L^2}
    \left(
      \frac{\pd \axy}{\pd y}
      + \frac{\axy}{1 - \tr \att / L^2}
        \frac{\pd \tr \att / L^2}{\pd y}
    \right)
    = 0.
\end{equation}
There are a few things to note about this system. The first
equation involves only $\axx$ and $\ayy$. Suppose now that
we can solve the constitutive equation~(\ref{eqn:fene_p_ce}) for
$\axx$ and $\ayy$; we can then always find a $p(y)$ to satisfy
the momentum balance. From the second equation we find that \emph{if}
$\axy \equiv 0$ is a solution of the constitutive equations (which
in uniform extensional flow it is, as we shall see), then this always
satisfies the momentum balance, similar to the UCM case. We can
therefore restrict ourselves to
finding a solution for the constitutive equations, given a uniform
extensional flow. An analytical solution is no longer possible,
and we will give numerical solutions. However, there is analytical
information to be obtained from the equations, mostly asymptotics.

Analogous to the UCM case, we assume that the extension depends only
on $y$, and we insert the pure extensional flow into the constitutive
equation~(\ref{eqn:fene_p_ce}):
\begin{equation}
  \begin{aligned}[]
    \Wi \left( -y A_{xx}'(y) -2 A_{xx}(y) \right)
                &= 
                -\left(
                  \frac{A_{xx}(y)}%
                    {1-\left(A_{xx}(y)+A_{yy}(y)\right)/L^2}
                    - \frac{1}{1-2/L^2}
                 \right), \\
    \Wi \left( -y A_{xy}'(y) \right)
                &= 
                -\frac{A_{xy}(y)}%
                    {1-\left(A_{xx}(y)+A_{yy}(y)\right)/L^2}, \\
    \Wi \left( -y A_{yy}'(y) + 2 A_{yy}(y) \right)
                &= 
                -\left(
                  \frac{A_{yy}(y)}%
                    {1-\left(A_{xx}(y)+A_{yy}(y)\right)/L^2}
                    - \frac{1}{1-2/L^2}
                 \right).
  \end{aligned}
  \label{eqn:fene_p_ext}
\end{equation}
We shall use the same domain and boundary conditions as for the UCM
case. For simplicity, we assume $\axx = \ayy = 1$
(the equilibrium values) at
$y = \pm 1$. It is clear that $A_{xy}\equiv 0$ is indeed a solution 
of the equations. We can then ignore this component of the
conformation tensor, and we may restrict ourselves to the diagonal
components.

\begin{figure}
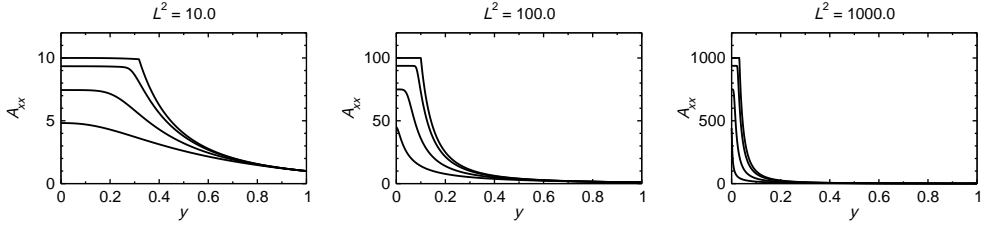

  \begin{center}
    \begin{tabular}{ccc}
      \includegraphics[width=0.29\textwidth]{strand10.eps} &
      \includegraphics[width=0.29\textwidth]{strand100.eps} &
      \includegraphics[width=0.29\textwidth]{strand1000.eps} 
    \end{tabular}
    \caption{Extension $\axx$ for different $L^2$, for $\Wi =
    0.9, 2.0, 8.0$, and $\Wi \to \infty$. Note the
    different scales on the vertical axis.\label{fig:conf_L2_1000_Wi} }
  \end{center}
\end{figure}

For $L^2 = 10, 100, 1000$ we then find, for different
$\Wi$, the plots in Fig.~\ref{fig:conf_L2_1000_Wi}. These are
clearly qualitatively similar to the birefringent strands that have
been found experimentally and numerically~\cite{mackley1975,crowley1976,fuller1980,harlen1992,feng1997,remmelgas1999}.

\begin{figure}
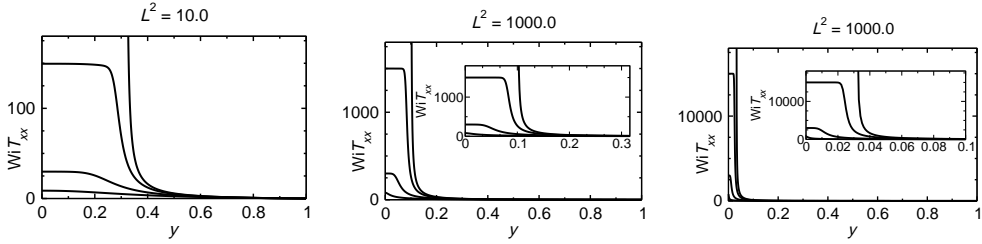

  \begin{center}
    \begin{tabular}{ccc}
      \includegraphics[width=0.29\textwidth]{stressxx10.eps} &
      \includegraphics[width=0.29\textwidth]{stressxx100inset.eps} &
      \includegraphics[width=0.29\textwidth]{stressxx1000inset.eps} 
    \end{tabular}
    \caption{Normal stress $\Wi\txx$ for different $L^2$, for $\Wi =
    0.9, 2.0, 8.0$ and $\Wi \to \infty$. Note the
    different scales on the vertical axis. The insets show a smaller
    range. \label{fig:stress_L2_1000_Wi} }
  \end{center}
\end{figure}

The formulation in terms of
a conformation tensor allows us to translate these results immediately
to a birefringence profile: for the relative change in index of
refraction $n$, we have the proportionality~\cite{peterlin1961}
\begin{equation}
  \frac{\Delta n}{n} \propto
    \sqrt{(\ayy - \axx)^2 + 4 \axy^2}.
\end{equation}
For the strongly stretched central region, the birefringence is
approximately directly proportional to $\axx$.

For the stresses, we can use Eq.~(\ref{eqn:stress_conformation}). Since
in our nondimensionalization, the \emph{physical} stress is given
by $\eta\,\Wi\,\tens{T}$, with $\eta$ the viscosity, we plot
$\Wi T_{xx}$ rather than $T_{xx}$ in Figure~\ref{fig:stress_L2_1000_Wi}.

\section{Asymptotic results for a FENE-P fluid}
We present four types of asymptotic results. The first is the
behaviour of the flanks of the strand. Here we shall find that we
recover the UCM  behaviour. Next, we look at the
maximal ``extension'' $\axx^{o}$ of the strand --- the
value of $\axx$ at $y = 0$. It depends on $\Wi$ and $L^2$.
We then combine these results, and give an approximate
expression for the width of the strand, which we define as the
point where the UCM profile intersects $\axx^{o}$.
We also show that this gives practically the same results as another
definition for the width, namely the inflection point of the
$\axx$ profile. Last, we look at the behaviour of the extension
around $y = 0$, and we show that even though the stresses at the
centre stay finite, stress \emph{gradients} may diverge for $y \to 0$.

\subsection{Outer flanks}

For small extensions, the FENE dumbbells behave approximately as
linear springs. We would therefore expect to recover UCM behaviour
outside the centre of the strand, where the extension is relatively
low.

To compare the FENE-P profiles to the UCM results, note that from
Eq.~(\ref{eqn:stress_conformation}) we find that for small extension
at large $L^2$
\begin{equation}
  \ttt \approx \frac{1}{\Wi} (\att - \itt)
    \quad \textrm{for} \quad \tr \att \ll L^2
    \quad \textrm{and} \quad L^2 \gg 1.
\end{equation}
For the $xx$ component, this implies
\begin{equation}
  \Wi \txx \approx \axx - 1.
\label{eqn:txx_axx}
\end{equation}
For the present boundary conditions, this means that outside the
centre of the strand, we have
\begin{equation}
  \axx \approx \Wi \txx^\textrm{UCM} + 1 =
     \frac{2\Wi}{1-2\Wi}
             \left(1 - |y|^{1/\Wi-2}\right) + 1.
\end{equation}
This approximation also works well for very high $\Wi$, because
the right hand side of Eq.~(\ref{eqn:fene_p_ce}) is then negligible,
until $\tr \att$ comes very close to $L^2$. Note that we 
compare the UCM \emph{stress} with the FENE-P \emph{extension}. For
the small extension limit, the comparison of the \emph{stresses} in
both models also gives good results (in that case, the
approximation~(\ref{eqn:txx_axx}) holds well), but it breaks down
in the large $\Wi$ limit.

\begin{figure}
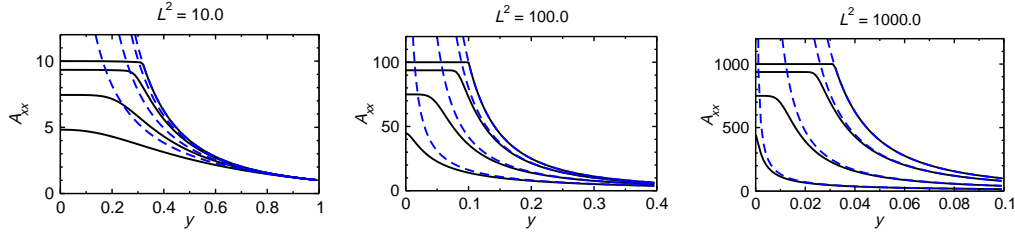

  \begin{center}
    \begin{tabular}{ccc}
      \includegraphics[width=0.3\textwidth]{cp10gr.eps} &
      \includegraphics[width=0.3\textwidth]{cp100gr.eps} &
      \includegraphics[width=0.3\textwidth]{cp1000gr.eps} 
    \end{tabular}
    \caption{Extension $\axx$, \emph{solid line}, compared to ``UCM extension''
    $\Wi \txx + 1$, \emph{dashed line}, for different $L^2$, for $\Wi =
    0.9, 2.0, 8.0$ and $\Wi \to \infty$. Note the different scales on both axes.\label{fig:axx_txx}}
  \end{center}
\end{figure}

If we compare the profiles for the conformation tensor to the
solution for the UCM fluid, as in Fig.~\ref{fig:axx_txx}, we see
that for Weissenberg number above
roughly $1.0$, the flanks of the stress profile are approximated by
those of the UCM case, cut off in the centre by the value of
$\axx$ at $y = 0$. This observation helps us to obtain some
asymptotic analytical results for the structure of the birefringent
strand.

\subsection{Maximal extension}
We can find the maximal extension by taking the limit of the
equations~(\ref{eqn:fene_p_ce})~and (\ref{eqn:fene_p_mb})
for $y \to 0$. The advection terms then vanish, and we are left
with the system
\begin{subequations}
\label{eqn:max}
\begin{align}
\label{eqn:axx_max}
    -2 \Wi \axx &=
      -\left(
         \frac{\axx}{1 - \tr \att / L^2}
         - \frac{1}{1 - 2 / L^2}
       \right), \\
\label{eqn:ayy_max}
    2 \Wi \ayy &=
      -\left(
         \frac{\ayy}{1 - \tr \att / L^2}
         - \frac{1}{1 - 2 / L^2}
       \right).
\end{align}
\end{subequations}
This system does not allow a simple analytical solution. However, we
can make a fair approximation by assuming that $\axx \gg
\ayy$ and $\axx \gg 1$ (this requires $L^2 \gg 1$). We
then have an autonomous equation for $\axx$:
\begin{equation}
  -2 \Wi \axx = -\left( \frac{\axx}{1 - \axx / L^2} \right).
\end{equation}
This is easily solved and gives
\begin{equation}
  \axx = L^2 \left(1 - \frac{1}{2 \Wi} \right).
  \label{eqn:axx_approx}
\end{equation}
This approximation can be shown to be self-consistent. For $\Wi \gtrsim
2$, clearly $\axx \gg 1$. Furthermore,
%equation~(\ref{eqn:ayy_max}) gives an upper bound for $\ayy$.
%We can rewrite this equation as
%\begin{equation}
%\label{eqn:ayy}
%  -c_1 \ayy + c_2 - 2 \Wi \ayy = 0
%  \quad \textrm{with} \quad
%  c_1 \equiv \frac{1}{1 - \tr \att / L^2}
%  \quad \textrm{and} \quad
%  c_2 \equiv \frac{1}{1 - 2 / L^2}.
%\end{equation}
%Noting that $\tr \att \geq 2$, regardless of the actual
%value of $\ayy$ (as long as it is positive), we find that
%\begin{equation}
%  c_1 \geq c_2 \geq 1.
%\end{equation}
%We can formally solve Eq.~(\ref{eqn:ayy}),
%\begin{equation}
%  \ayy = \frac{c_2}{c_1 + 2 \Wi},
%\end{equation}
%and with the given inequality we find that
%\begin{equation}
%  \ayy \leq 1.
%\end{equation}
since the flow is purely compressive in the $y$-direction, we expect
$A_{yy} \leq 1$. In fact, we can find an asymptotic expression for
$A_{yy}$ at $y = 0$ from Eq.~(\ref{eqn:ayy_max}). For large $L^2$,
the rightmost term is approximately 1, and assuming
$A_{yy} \ll A_{xx}$, the equation simplifies to 
\begin{equation}
  2 \Wi A_{yy} - \frac{A_{yy}}{1+A_{xx}/L^2} = 1.
\end{equation}
Inserting the asymptotic expression~(\ref{eqn:axx_approx}) for
$A_{xx}$, we find
\begin{equation}
\label{eqn:ayy_approx}
  A_{yy} = \frac{1}{4 \Wi}.
\end{equation}
For large $L$ and $\Wi \gtrsim 2$, this implies that
$A_{yy}$ is of $O(1/L^2)$ compared to $A_{xx}$, and the above
approximation is self-consistent. We conclude that for $L^2 \gg 1$
and $\Wi \gtrsim 2$, Eq.~(\ref{eqn:axx_approx}) is a reasonable
approximation.

\begin{figure}
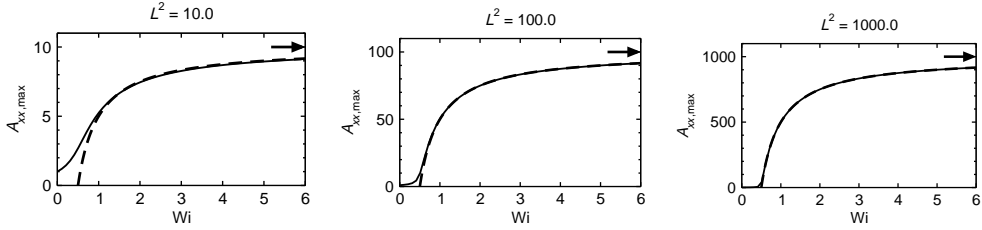

  \begin{center}
    \begin{tabular}{ccc}
      \includegraphics[width=0.29\textwidth]{max10gr.eps} &
      \includegraphics[width=0.29\textwidth]{max100gr.eps} &
      \includegraphics[width=0.29\textwidth]{max1000gr.eps} 
    \end{tabular}
    \caption{Maximum of $\axx$ as a function of $\Wi$ for different
             $L^2$. Numerical solution, \emph{solid line},
             compared with the approximation~(\ref{eqn:axx_approx}),
             \emph{dashed line}. The arrow indicates the asymptotic value
             $L^2$ for $\Wi \to \infty$. Note the different scales on
             the vertical axis. \label{fig:max_num}}
  \end{center}
\end{figure}

This is confirmed by numerical solution of the system~(\ref{eqn:max}).
In Fig.~\ref{fig:max_num} we show the maximum value of $\axx$ as a
function of $\Wi$ for different $L^2$.

\subsection{Width of the birefringent strand}
Given these results, we can now proceed to derive an approximation
for the width of the birefringent strand as a function of $\Wi$ and
$L^2$. We define the width as the location where the UCM approximation
intersects the value of the ``centre maximum'' of the extension.
This is essentially the width of the ``plateau'' in
Fig.~\ref{fig:conf_L2_1000_Wi}.

\begin{figure}
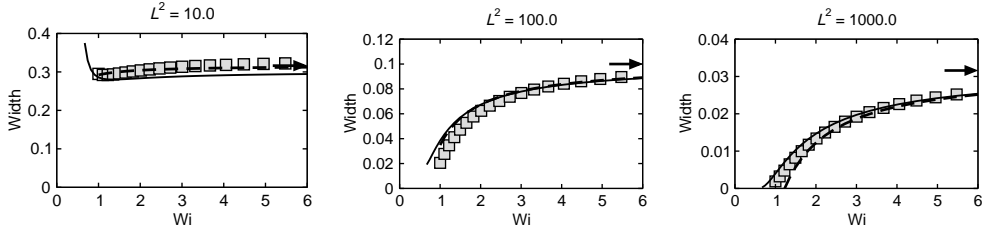

  \begin{center}
    \begin{tabular}{ccc}
      \includegraphics[width=0.29\textwidth]{width10gr.eps} &
      \includegraphics[width=0.29\textwidth]{width100gr.eps} &
      \includegraphics[width=0.29\textwidth]{width1000gr.eps} 
    \end{tabular}
    \caption{Width of the birefringent strand as a function of
             $\Wi$ for different $L^2$. ``Exact'' solution,
             Eq.~(\ref{eqn:width_exact}), \emph{solid line},
             compared with the approximation~(\ref{eqn:width_approx}),
             \emph{dashed line}. The arrow indicates the
             asymptotic value $1 / L$, and the \emph{squares}
             denote a numerical result where the width is defined
             by the inflection point of the stress profile. Note the
             different scales on the vertical axis. \label{fig:width}}
  \end{center}
\end{figure}

We combine the results in eqs.~(\ref{eqn:ucm_sol}), 
(\ref{eqn:txx_axx}) and (\ref{eqn:axx_approx}) to obtain for the
intersection point $y_0$
\begin{equation}
  \frac{2 \Wi}{1 - 2 \Wi} \left(1 - | y_0 |^{1 / \Wi - 2}\right) + 1
  =
  L^2 \left( 1 - \frac{1}{2 \Wi} \right).
\label{eqn:width_eqn}
\end{equation}
Solving for $y_0$ thus gives the strand half-width as
\begin{equation}
  y_0 = \left[
          \frac{1}{2 \Wi} + L^2 \left(1 - \frac{1}{2 \Wi}\right)^2
        \right]^{\Wi / (1 - 2 \Wi)}.
\label{eqn:width_exact}
\end{equation}
For $\Wi \to \infty$ we find $y_0 \to 1 / L$. An expansion around
$\Wi = \infty$ gives
\begin{equation}
  y_0 \approx \frac{1}{L}
            + \frac{2 L^2 (1 - \ln L) - 1}{4 L^3} \frac{1}{\Wi}
            + O\left(\frac{1}{\Wi^2}\right).
\end{equation}
For $L^2 \gg 1$ this reduces to
\begin{equation}
  y_0 \approx \frac{1}{L}
            + \frac{1 - \ln L}{2 L} \frac{1}{\Wi}
            + O\left(\frac{1}{\Wi^2}\right).
\label{eqn:width_approx}
\end{equation}
The strand half-width in these approximations is shown in
Fig.~\ref{fig:width} for different values of $L^2$. Note that
the ``exact'' solution~(\ref{eqn:width_exact}) bends up for lower
$\Wi$. This is unphysical. It shows up because both sides
of Eq.~(\ref{eqn:width_eqn}) are approximations that break down
for relatively small $L^2$ and $\Wi$.

The asymptotic value $1/L$ has already been found by Mackley and
coworkers~\cite{crowley1976}, in a more qualitative way.

\subsection{Central region}
The behaviour of the extension (and, hence, the normal stresses)
as $y \to 0$, is not immediately clear from the plots in
Figures~\ref{fig:conf_L2_1000_Wi} and \ref{fig:stress_L2_1000_Wi}.
In particular, the question arises
whether the stresses display singular behaviour for $y \to 0$, as
was found by Renardy for the Giesekus model~\cite{renardy2006}.

It is clear from Eq.~(\ref{eqn:fene_p_ext})
that $y=0$ is a singular point of this system of differential
equations: as $y \to 0$, the higest-order derivative vanishes. The
conformation tensor will therefore not be analytical
in $y$ at $y=0$. To determine the degree of this singularity, we 
follow the usual procedure for dealing with singular
points~\cite{benderorszag1978}, and we expand the solution around
$y=0$ as
\begin{equation}
  \tens{A}(y) = \sum_{r = 0}^{\infty} \tens{a}_r \, y^{\beta + r}
                + \tens{A}_0,
\label{eqn:singular_series}
\end{equation}
with $\beta$ in general non-integer, and $\tens{A}_0$ the uniform
solution of Eqs.~(\ref{eqn:max}). We then seek an expression for the
exponent $\beta$ of the dominant (lowest-order) term by balancing the
leading-order singular terms. For the small-extension limit, the
equations actually reduce to the UCM equations, and the exponent is
given by the exact solution Eq.~(\ref{eqn:ucm_sol}), that is,
\begin{equation}
  \beta = \frac{1}{\Wi}-2 \qquad \textrm{for} \quad\tr\tens{A}\ll L^2.
  \label{eqn:asymp1}
\end{equation}

For moderate to high extension, we need to take into account the
finite extensibility of the polymers. To lowest nontrivial order in
$y$, the expansion~(\ref{eqn:singular_series}) becomes, in components,
\begin{equation}
  \begin{aligned}[]
    A_{xx}(y) &=& A_{xx}^{o} + a_{xx} y^\beta,\\
    A_{yy}(y) &=& A_{yy}^{o} + a_{yy} y^\beta.
  \end{aligned}
\end{equation}

We insert this into the constitutive equations~(\ref{eqn:fene_p_ce})
for purely extensional flow, and expand the fractions in that
equation to first order in $a_{xx}$ and $a_{yy}$. We then have at first
order, leaving out the constant terms (which form the equation for
$\tens{A}_0$),
\begin{equation}
  \begin{aligned}[]
      \frac{a_{xx}}{1-\tr \att_0 / L^2} 
      + \frac{A_{xx}^{o}(a_{xx}+a_{yy}) / L^2}{(1 - \tr \att_0 / L^2)^2}
      - 2 \Wi a_{xx} - \beta \Wi a_{xx}
    &=& 0, \\
      \frac{a_{yy}}{1-\tr \att_0 / L^2} 
      + \frac{A_{yy}^{o}(a_{xx}+a_{yy}) / L^2}{(1 - \tr \att_0 / L^2)^2}
      + 2 \Wi a_{yy} - \beta \Wi a_{yy}
    &=& 0.
  \end{aligned}
\end{equation}

We can simplify these equations considerably. We
put $A_{yy}^{o} \to 0$, because according to~(\ref{eqn:ayy_max}) 
it is of order $1/L^2$ compared to the other terms and also
compared to $1-A_{xx}^{o}/L^2$. We then substitute the asymptotic
result~(\ref{eqn:axx_approx}) for $A_{xx}^{o}$. Since the
system is linear in $a_{xx}$ and $a_{yy}$, we can scale them such that
\begin{equation}
  a_{xx} = 1 \qquad \textrm{and} \qquad a_{yy} = \gamma,
\end{equation}
where $\gamma$ is the ratio $a_{yy} / a_{xx}$. We are then left with
the system
\begin{equation}
  \begin{aligned}
    \Wi [ (4 \Wi - 2)(1 + \gamma) - \beta ] = 0, \\
    \Wi \gamma (\beta - 4) = 0,
  \end{aligned}
\end{equation}
which has two solutions:
\begin{equation}
  \gamma = 0, \quad \beta = 4\Wi -2 \qquad \textrm{and} \qquad
  \gamma = -\frac{2 \Wi - 3}{2 \Wi - 1}, \quad \beta = 4.
  \label{eqn:asymp23}
\end{equation}

\begin{figure}
  \begin{center}
    \includegraphics[width=0.6\textwidth]{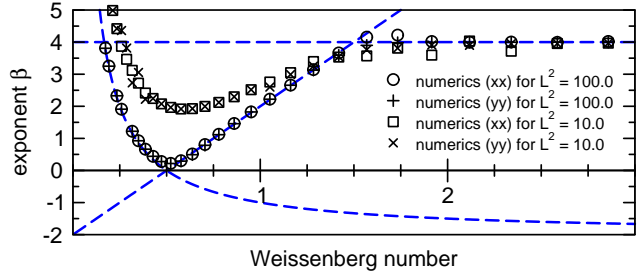} 
    \caption{The exponent $\beta$ as a function of $\Wi$. The dashed
    lines are the asymptotic approximations: the curve is from
    Eq.~(\ref{eqn:asymp1}), the two straight lines are from
    Eq.~(\ref{eqn:asymp23}). The dots were calculated from numerical
    solutions for $a_{xx}$ and $a_{yy}$.
\label{fig:exponent}}
  \end{center}
\end{figure}

The three asymptotic results for $\beta$ in Eqs.~(\ref{eqn:asymp1})
and~(\ref{eqn:asymp23}) are plotted in Figure~\ref{fig:exponent},
together with numerical results for $L^2 = 10.0$ and $L^2 = 100.0$.
These were obtained by integrating the differential
equations~(\ref{eqn:fene_p_ext})
numerically from $y = 10^{-10}$ to $y = 1.0$ and taking the slope in
a log-log representation of $a_{xx}$ and $a_{yy}$. Because the stress
is a regular function of the extension, these exponents immediately
carry over to the stresses around $y=0$.

From the asymptotic results, we can conclude the following: although
the stresses themselves remain
finite for any Weissenberg number, for sufficiently large $L$ there
is a range of $\Wi$ for
which $\beta < 1$ and stress gradients become infinite. This range
lies roughly between $\Wi = 1/3$ and $\Wi = 3/4$. These bounds are
obtained by putting $\beta = 1$ in Eqs.~(\ref{eqn:asymp1})
and~(\ref{eqn:asymp23}). The range is finite, because for low $\Wi$
the velocity gradient is insufficient to cause large extension
gradients, while for high $\Wi$ the polymers are already almost
fully stretched well outside of the central region, which also
prevents the formation of large extension gradients in the central
region (as can be seen in Fig.~\ref{fig:axx_txx}).

The precise extent of the range in $\Wi$ for which the stress
gradient diverges, depends on $L^2$; for small $L^2$ it is absent
entirely, as can be seen in Fig.~\ref{fig:exponent} for $L^2 = 10.0$.
Although the analysis in the present paper is less straightforward,
it is fully analogous to Renardy's result for the Giesekus
model~\cite{renardy2006}.

\section{Concluding remarks}
The results that we have derived here are valid for a rather artificial
geometry, namely a purely extensional flow on an infinite strip.
However, we believe that the results are still relevant for more 
realistic flows. The approximation that quantities do not depend on $x$
is valid on the central incoming flow line. If we can show that the
results do not depend crucially on
the exact boundary conditions and on the assumption of uniform
extensional flow, then the results that were derived above still give
information on the shape of actual birefringent strands in fluids
that are described by the FENE-P model.

We hope that our results for the scaling behaviour of these
near-singular structures will stimulate new experiments on the
birefringence strands --- we surmise that the birefringent strands
seen experimentally are indeed the experimental realizations of
these structures, but the data on these in the existing literature
lack the precision to test this claim. 

Finally, we make a connection between our results and the recently
observed instabilities in the cross-channel flow \cite{arratia2006}
(see Fig.~\ref{fig:int_stag_flow}b). There, the fully-developed
(or being close to it) Poiseuille flow in the inlet channels provides
boundary conditions for the normal and shear components of the
stress tensor ($T_{yy}$ and $T_{xy}$ in our notation) at the inflow
boundaries of the square region in the centre of the flow domain.
These boundary conditions clearly differ from the special values
discussed in Section 2, and thus the flow in the central region
is \emph{not} a constant-stress solution as
Eq.~(\ref{eqn:unif_sol_ucm}) and is expected to be dominated by
strands similar to (\ref{eqn:ucm_sol}). Moreover, the non-zero value
of the shear stress and the parabolic profile of the velocity at the
boundaries imply that the actual flow in the central domain is a
combination of elongational and shear components. One might then
argue that the first instability observed in~\cite{arratia2006}
corresponds to switching from elongation-dominated to shear-dominated
velocity field, while the second bifurcation would be a purely
elastic instability of that shear-dominated flow with curved
streamlines \cite{larson1992,morozov2007}. The latter are almost
always Hopf bifurcations \cite{larson1992,morozov2007} which is
consistent with the time-dependent flows observed
in~\cite{arratia2006} at large Weissenberg numbers. Our results can
be considered to be a first step in constructing an analytic
approximation to the base flow in the cross-channel geometry which
can then be used in linear stability analysis. Inclusion of finite
extensibility is likely to be important in studies of the secondary
instability.

\bibliography{singular}
  \bibliographystyle{elsart-num}

\end{document}